\documentclass[aps,prl,twocolumn,superscriptaddress,showpacs]{revtex4}
\usepackage[T1]{fontenc}
\usepackage{graphicx}

\begin{document}

\title{Triboelectric charging of volcanic ash from the 2011 Gr\'{i}msv\"{o}tn eruption}

\author{Isobel M. P. Houghton}
\email[]{isobel.houghton@bristol.ac.uk}
\affiliation{Department of Electrical and Electronic Engineering, University of Bristol, Bristol, BS8 1UB, UK}
\affiliation{Centre for Nanoscience and Quantum Information, University of Bristol, Bristol, BS8 1FD, UK}
\author{Karen L. Aplin}
\email[]{k.aplin1@physics.ox.ac.uk}
\affiliation{Department of Physics, University of Oxford, Oxford, OX1 3RH, UK}
\author{Keri A. Nicoll}
\email[]{k.a.nicoll@reading.ac.uk}
\affiliation{Department of Meteorology, University of Reading, Reading, RG6 6BB, UK}

\date{\today}

\begin{abstract}
The plume from the 2011 eruption of Gr\'{i}msv\"{o}tn was highly electrically charged, as shown by the considerable lightning activity measured by the UK Met Office's low-frequency lightning detection network. Previous measurements of volcanic plumes have shown that ash particles are electrically charged up to hundreds of km away from the vent, which indicates that the ash continues to charge in the plume [Harrison \emph{et al.}, \emph{Env. Res. Lett.} \textbf{5} 024004 (2010), Hatakeyama \emph{J. Met. Soc. Japan} \textbf{27} 372 (1949)]. In this paper we study triboelectric charging of different size fractions of a sample of volcanic ash experimentally. Consistently with previous work, we find that the particle size distribution is a determining factor in the charging. Specifically, our laboratory experiments demonstrate that the normalised span of the particle size distribution plays an important role in the magnitude of charging generated. The influence of the normalised span on plume charging suggests that all ash plumes are likely to be charged, with implications for remote sensing and plume lifetime through scavenging effects. 

\end{abstract}


\maketitle

Volcanic ash is known to charge electrically, producing some of the most spectacular displays of lightning on the planet~\cite{Mather,James}.  Lightning activity within volcanic plumes can be sensed remotely using systems such as the UK Met Office long-range lightning detection network, ATDnet \,\cite{Bennett}, which recorded over 16\,000 lightning strokes during the 2011 Gr\'{i}msv\"{o}tn eruption~\cite{Arason}. These remote sensing techniques can only be fully exploited if the charging mechanisms in volcanic plumes are well understood. Although the exact details of ash charging processes will vary from one eruption to another, triboelectrification, fractoemission and the `dirty thunderstorm' mechanism~\cite{Mather,James,Williams} are all thought to play a role in the electrification of ash near the vent.  In addition to near-vent charging, observations show that charging can also occur in volcanic plumes up to hundreds of km from the source region~\cite{Harrison, Hatakeyama, Arason2}.  The sustained nature of this charge in the presence of electrically conducting air, suggests that a self-charging mechanism through the action of ash-to-ash contact charging (triboelectrification), may also play a role in the electrification of volcanic ash.  Previous theoretical work on triboelectric charging of single-material particle systems has shown that the charging is determined by the number size distribution~\cite{Lacks}.  This paper details a laboratory investigation into triboelectric charging of a sample of ash from the Gr\'{i}msv\"{o}tn  eruption in Iceland in 2011, in terms of the particle size distribution, using specially designed apparatus.

Charging arising from contact between two different material surfaces can be understood as a result of the different work functions of the materials, however triboelectric charging in systems of identical materials cannot be explained in this way. Lowell and Truscott presented a model for triboelectric charging between macroscopic samples of identical materials based on spatial localisation of electrons on the material surface~\cite{Lowell}. Spatial localisation of electrons prevents relaxation of electrons in high energy states to vacant low energy states elsewhere in the material. Contact between two surfaces provides a relaxation mechanism where a localised high energy electron on one surface can move to a vacant low energy state on the other surface, resulting in electron transfer between surfaces. 

This model has more recently been developed to describe triboelectric charging of granular systems~\cite{Lacks}. The number of trapped high energy electrons is assumed to be proportional to the particle's surface area, \emph{i.e.} the surface charge density is the same for all particles, and the number of low energy electrons is zero. In a collision, a high energy electron in one particle will be transferred to a low energy state in the other particle. If both particles have equal numbers of high energy electrons, there is no net charge transfer. However, if only one particle has a high energy electron, this will be lost to the other particle. Smaller particles will therefore lose all their trapped high energy electrons before the larger particles, while continuing to receive electrons into vacant low energy states, causing net electron transfer from large to small particles. This results in an average negative charge on the smaller particles and average positive charge on the larger particles. 

Lacks and Levandovsky present simulations of particle dynamics of simple granular systems to illustrate their model, which reproduces the negative charging of smaller particles and positive charging of larger particles observed empirically~\cite{Lacks}. This model has since been extended to include geometric considerations which favour electron tunnelling from large particles to small~\cite{Kok}.  In addition to these numerical studies, experimental studies of triboelectric charging in soda lime glass, Mars and lunar regolith simulants  demonstrate that the particle size dependent charging seen in natural phenomena (e.g. dust devils, volcanic plumes) can be reproduced in the laboratory~\cite{Forward1, Forward2, Forward3, Krauss}. 

P\"{a}htz \emph{et al.} present a model of electron transfer between identical dielectric grains in an electric field~\cite{Pahtz}. The applied field polarises the grains and when two oppositely charged surfaces collide, electrons are transferred. Following separation, the applied field repolarises the grains. This charging model is not applicable to our experiments as care is taken to ensure there are no applied external electric fields, however it may contribute to charging in plumes. Despite this recent progress, the details of particle charging as a function of size distribution are not well understood.

\begin{figure}
\includegraphics[width=0.9\columnwidth]{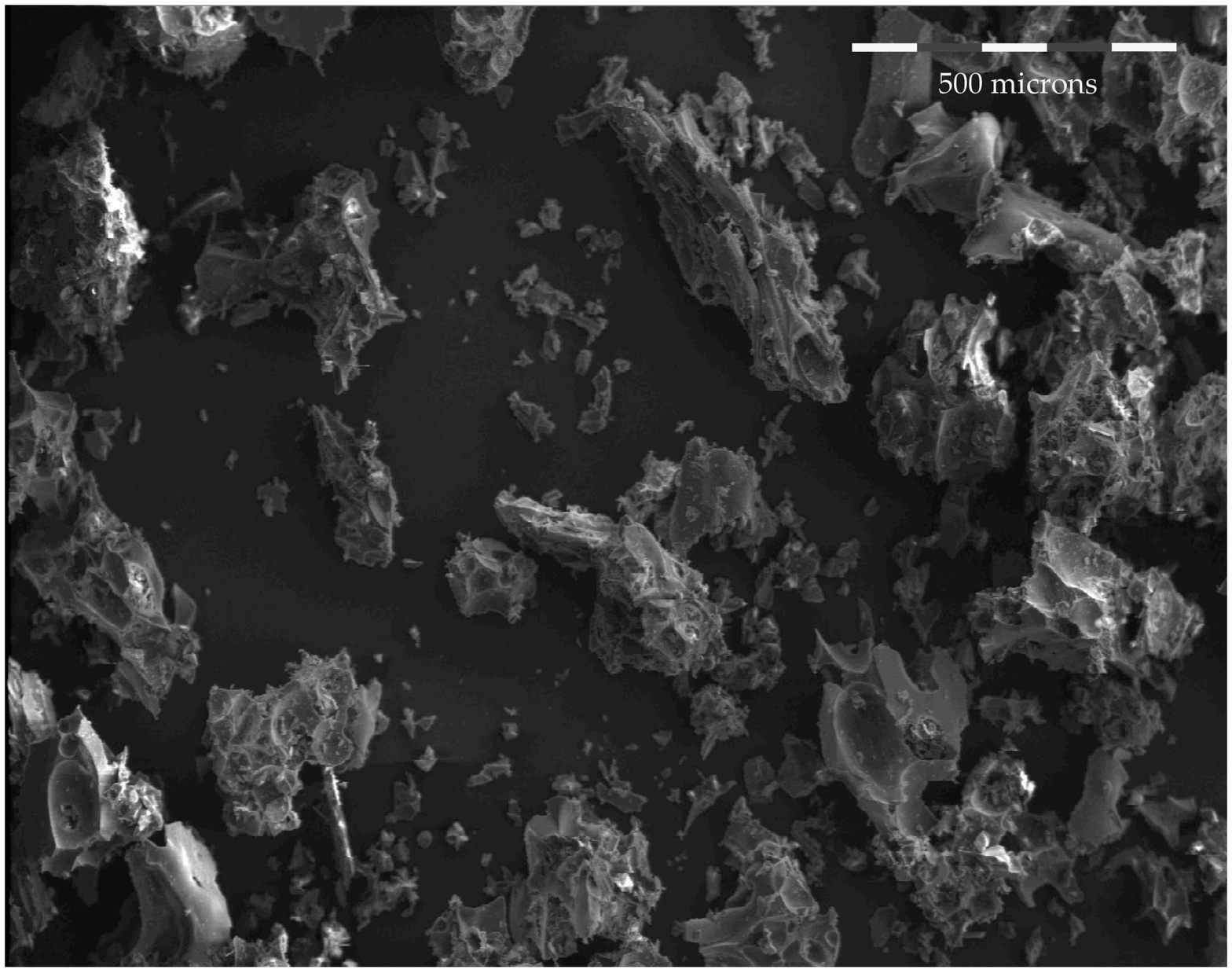} 
\caption{\label{Fig_1}Scanning electron microscope image of ash sample. The non-spherical nature of the particles is clear. Image courtesy of David Pyle.}
\end{figure}

The 2011 eruption of the Gr\'{i}msv\"{o}tn volcano began on 21 May, and the eruption was associated with considerable volcanic lightning~\cite{Arason}. Preliminary estimates have put the total amount of tephra ejected at 0.6-0.8\,km$^3$ dense rock equivalent (DRE)~\cite{Gudmundsson}. Ash was collected on 26 May at Kirkjub\ae jarklaustur, approximately 75\,km south-south-west of the Gr\'{i}msv\"{o}tn crater. The eruption ended on 28 May. Scanning electron microscope images of the sample (Figure~\ref{Fig_1}) show the particles to be angular and to have a wide range of sizes. 

\begin{figure}
\includegraphics[width=0.9\columnwidth]{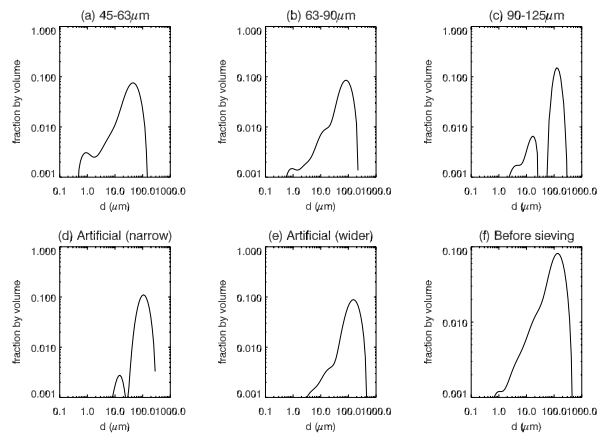}
\caption{\label{Fig_2}Volumetric particle size distributions, all measured with the Malvern Mastersizer 2000. (a)-(c) sieved samples, with the size fractions defined by the sieves indicated (d) artificial narrow bimodal distribution (50:50 mixture of 45-63\,$\mu$m and 90-125\,$\mu$m) (e) artificial wider bimodal distribution (50:50 mixture of 45-63\,$\mu$m and 125-180\,$\mu$m) and (f) the size distribution of the sample before sieving.}
\end{figure}

Ash diameter distributions were measured with a Malvern Mastersizer 2000, which uses laser diffraction to calculate volumetric size distributions of suspended samples, in the range 0.02-2000 $\mu$m to better than 1\%~\cite{Malvern}. Volumetric size distributions were obtained  both before and after the samples were separated into different size fractions by geological sieving (dry, rather than wet, sieving was used to preserve aggregates that could contribute to the plume's electrostatic properties~\cite{Green}), as shown in Figure~\ref{Fig_2} and summarised in Table~\ref{size_dist_table}.  Dry sieving showed that the larger ash particles were slightly darker in colour than the smaller particles. This variation in optical properties suggests the ash sample is made of a mixture of different substances, consistent with other observations~\cite{Kerminen}. These different substances may triboelectrically interact with each other, in addition to the charge transfer as a function of size. 

Three sieved samples were tested, from nominally 45-63\,$\mu$m, 63-90\,$\mu$m and 90-125\,$\mu$m size distributions, defined by the sieves. Two artificial size distributions were created from mixing 50:50 samples (by mass) of 45-63\,$\mu$m and 90-125\,$\mu$m, and 45-63\,$\mu$m  and 125-180\,$\mu$m, to generate a narrow bimodal and a wide bimodal distribution. 

\begin{table}
\begin{ruledtabular}
\begin{tabular}{| c | c | c | c |}
Size distribution & Normalised & Modality &Distribution\\
 & span & coefficient & group\\
\hline
\hline
Before sieving  &1.919 & 0.832 & - \\
\hline
45-63\,$\mu$m & 1.801 & 0.859 &A\\ 
63-90\,$\mu$m  & 1.642 & 0.872 & B\\
90-125\,$\mu$m  & 0.852 & 0.906 & C\\
125-180\,$\mu$m  & 0.775 & 0.909 & - \\ 
\hline
45-63 and 90-125\,$\mu$m  & 1.144 & 0.892 & C\\
45-63 and 125-180\,$\mu$m  & 1.469 & 0.877 & B\\
\end{tabular}
\end{ruledtabular}
\caption{\label{size_dist_table}
Volumetric size distribution summaries. The normalised span is a non-dimensional index of the polydispersity of the distribution, defined by the normalised interdecile range~\cite{Malvern}. The modality coefficient \emph{b} is calculated from the skewness and kurtosis of the distribution~\cite{SAS}. The five samples tested are divided into three groups: A, B and C. Distribution A has a broad span and is less bimodal (span~$>1.3$, $b<0.86$), distribution B has a broad span and is more bimodal (span~$>1.3$, $b>0.86$) and distribution C has a narrow span and is more bimodal (span~$<1.3$, $b>0.86$). }
\end{table}

Here we use the term \emph{normalised span}, a non-dimensional index of the polydispersity of the distribution, defined as the difference between the $90^{th}$ and $10^{th}$ diameter percentiles, divided by the median diameter~\cite{Malvern}. The pre-sieve size distribution contained particles between 1-500$\mu$m in diameter, with most of the particles between 20 and 200\,$\mu$m. The sieve and Malvern sizer diameters only agree approximately, which may be due to the assumption of sphericity used by the Malvern instrument; this is clearly incorrect  as demonstrated by Figure~\ref{Fig_1}.  The sieved samples' size distributions all show a substantial tail of fine particles, which we believe cannot be from cross-contamination within the particle sizer, due to careful experimental technique. An alternative explanation for the fine tail could be disaggregation whilst the samples were in the Malvern sizer, however, a similar fine tail would be expected across all the measurements, which was not seen (Figure~\ref{Fig_2}). The fall speed of these fine particles is small in relation to the larger particles, therefore they are not thought to contribute substantially to the results reported. 

To quantify the modality of the samples, the modality coefficient \emph{b} was calculated using $b = \frac{\gamma^2+1}{\kappa}$~\cite{SAS}, where $\gamma$ is the skewness and $\kappa$ the kurtosis of the distribution, determined using the scientific programming language IDL~\cite{IDL}. As \emph{b} increases, the distribution becomes more bimodal~\cite{SAS}. Using the modality coefficient and span allows the samples to be divided into three groups (A, B and C) as indicated in Table~\ref{size_dist_table}. 

\begin{figure}
\includegraphics[width=0.9\columnwidth]{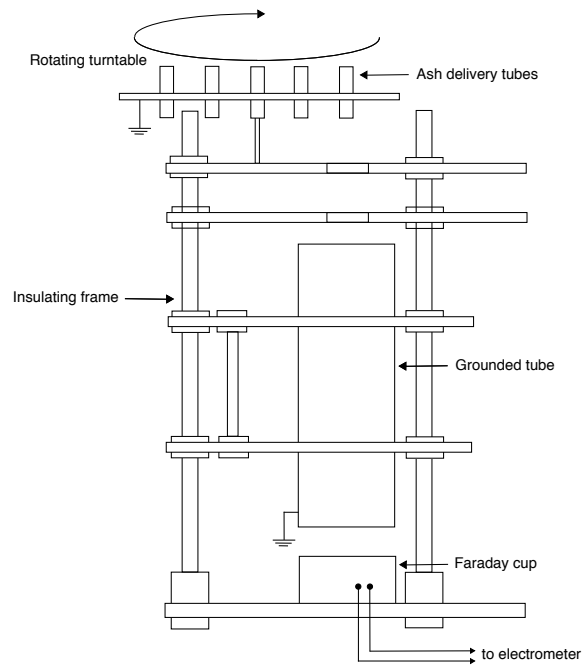}
\caption{\label{Fig_3}Schematic diagram of the ash charge apparatus showing the ash delivery apparatus at the top and collecting Faraday cup at the base. The grounded tube is 80\,mm in diameter and 310\,mm long. The base of the grounded tube is 40\,mm above the top of the Faraday cup. The delivery tubes are 10\,mm in diameter and are 450\,mm above the top of the Faraday cup.}
\end{figure}

Electrostatic charging of the Gr\'{i}msv\"{o}tn ash was investigated using a grounded tube, supported by an insulating frame, through which ash is dropped into an isolated Faraday cup connected to a sensitive electrometer, shown in Figure~\ref{Fig_3}. In an optimised and consistent delivery technique, ash was delivered to the charge apparatus via identical individual release mechanisms mounted on a grounded rotating metal turntable, supported vertically above the inlet. The charge, $\Delta Q$, transferred from the ash to the Faraday cup is related to the change in voltage, $\Delta V$ measured at the cup and the capacitance, $C$, of the system (130\,pF), by the relationship $\Delta Q = C \Delta V$. The change in Faraday cup voltage was recorded by a Campbell CR3000 data logger at $300\,$Hz.

For each ash charging experiment, 0.2\,g of Gr\'{i}msv\"{o}tn ash (baked to remove water) was weighed and transferred to an individual delivery tube.  To minimise the unwanted effect of self-charging of the ash during handling, ash was left in the grounded delivery tube for 30\,minutes before each test. This would allow time for any residual charge on the ash to decay, assuming that air has an electrical conductivity of $10^{-15}\,$S\,m$^{-1}$ \cite{Aplin}.  We therefore assume that any charge measured on the ash after descent is entirely from ash-to-ash contact (triboelectric) charging during interactions whilst the ash is falling under gravity, analogously to a volcanic plume in the atmosphere. During charging experiments, ash was observed to fall in a narrow column of similar dimensions to the ash delivery tube, meaning that interactions between ash and the walls of the grounded tube are unlikely. As the apparatus is grounded, we believe that no other charging processes can have a significant effect on the ash sample.

Charging experiments were undertaken with the five different size distributions of Gr\'{i}msv\"{o}tn ash described above: the three sieved fractions and the two artificial distributions.  For each size distribution, ten ash charging experiments were performed to reduce the sampling error.

\begin{figure}
\includegraphics[width=0.9\columnwidth]{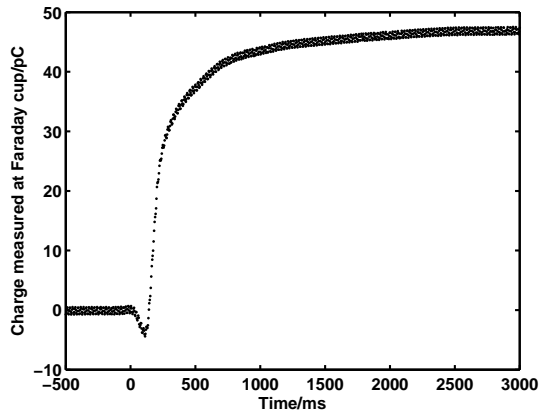}
\caption{\label{Fig_4} A typical Faraday cup charge measurement trace. The charge difference between the initial value and asymptotic final value are used to compare the different experiments.  The net charge differences are summarised in the box plot shown in Figure~\ref{Fig_5}. }
\end{figure}

\begin{figure}
\includegraphics[width=0.9\columnwidth]{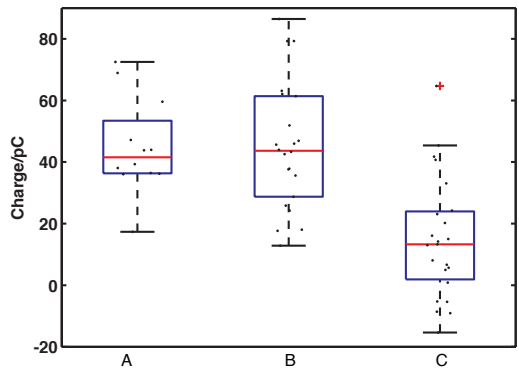}
\caption{\label{Fig_5}  Net charge difference at the Faraday cup for the three groups of samples: A (broad span, less bimodal), B (broad span, more bimodal) and C samples (narrow span, more bimodal). The central mark in each box (red, color online) shows the median change for each group. The edges of the box show the $25^{th}$ and $75^{th}$ percentiles and the whiskers extend to 1.5 times the inter-quartile range. Data points outside this limit are shown as (red) crosses. }
\end{figure}

Figure~\ref{Fig_4} shows a typical charging trace measured at the Faraday cup. The ash is released at $t = 0$ and falls with a typical speed of 1\,m\,s$^{-1}$. Initially the charge decreases to a minimum value before increasing to a maximum value. The change in charge measured by the Faraday cup for the three types of samples are summarized as box and whisker plots in Figure~\ref{Fig_5}, and their distributions compared using the Wilcoxon test, using a critical value of 5\%.

Figure~\ref{Fig_5} shows the net charge difference for all three groups.  When comparing B and C, distribution C shows much smaller charging (median value 13.3 pC) than distribution B (median value 43.7 pC) with a 95\% confidence interval ($p-$value $< 0.05$). Comparing both broad span distribution groups (A and B) against C does not alter this result.

For A and B there is no difference, again with a 95\,\% confidence interval, between samples with more or less bimodal distribution: the median values of the charge difference are $41.6$\,pC (A) and $43.7$\,pC (B). This demonstrates that the normalised span of the distribution may affect the magnitude of the charging, while the modality of the sample does not. The results obtained by comparing samples grouped in terms of span and modality rather than size also indicates that the particle distribution has a greater effect on charging than any effects of varying composition with size.

In conclusion, charging experiments show that Gr\'{i}msv\"{o}tn ash is easily electrified via the self-charging mechanism, with the span of the particle size distribution playing an important role in the magnitude of the charge generated.  Samples with the largest normalised span in particle sizes (\emph{i.e.}~a variety of different sized particles) were observed to generate the largest magnitude charges.  This agrees with the laboratory findings of Krauss \emph{et al.}, who also found increased charging with a broad particle-size distribution during experiments with Martian regolith stimulant~\cite{Krauss}. It is also observed that the span of the particle distribution dominates over modality.  

Charged aerosol particles, such as those found by Harrison \emph{et al.} in the 2010 Ejyafjallaj\"{o}kull plume, are preferentially removed (scavenged), by water droplets~\cite{Harrison}. For example, Tinsley \emph{et al.} (2000)~\cite{Tinsley} and Tripathi and Harrison (2002)~\cite{Tripathi} both show that the collision efficiencies of  particles between 1-10$\mu$m diameter with water droplets are enhanced by a factor of 30, even with relatively few charges ($50\bar{e}$) on the particle. 

Our experiments on ash samples with a wide size distribution of between 1-500$\mu$m demonstrated that the samples charged according to the span of the size distribution. All volcanic plumes will therefore self-charge triboelectrically to some extent, and will contain a fine tail of charged particles, as observed, that will affect scavenging. Preferential removal of small charged particles is likely to shorten the plume lifetime, particularly  at locations distant from the vent where the larger particles have already been lost.

These findings have implications for the remote sensing of volcanic ash via electrostatic techniques as the amount of charging will change with the particle size distribution, giving different charging behaviour in different eruptions, in different phases of an eruption and as the particle size distribution changes through gravitational settling. We also expect the triboelectric charging of volcanic plumes to be relevant in planetary atmospheres. Sustained triboelectric self-charging of volcanic plumes distant from the vent is possible as long as there is a distinct particle size distribution. Other mechanisms may also act to enhance the charging~\cite{James,Williams,Pahtz}.

The authors would like to thank Dr.~\TH\'{o}r\dh ur Arason and Krist\'{i}n Hermannsd\'{o}ttir (Icelandic Meteorological Office) for providing the sample; Prof.~David Pyle, Dr.~Tamsin Mather and Owen Green (Earth Sciences Department, University of Oxford) for their assistance in characterising the sample and Dr.~Mona Edwards (Geography Department, University of Oxford) for access to the Malvern Mastersizer 2000. Technical support was provided by Keith Long (Physics Department, University of Oxford), and the original ash charging apparatus was constructed by technical staff from the Department of Meteorology at the University of Reading. KAN acknowledges the support of the Leverhulme Trust through an Early Career Fellowship. This work received support through a small grant from the Cabot Institute, University of Bristol.

\bibliography{Grimsvotn_bib}

\end{document}